\documentclass[a4paper]{article}

\usepackage{icrc2013}
\usepackage[english]{babel}

\usepackage{amsmath}
\usepackage{amssymb}
\usepackage{subscript}

\usepackage[separate-uncertainty=true,
 range-units = single,
 range-phrase = -
]{siunitx}
\newcommand{\EeV}{\exa\electronvolt}
\DeclareSIUnit\parsec{pc}
\DeclareSIUnit\gauss{G}

\newcommand{\vecN}[1]{\mathbf{#1}} 
 
\title{Detecting Local Deflection Patterns of Ultra-high Energy Cosmic
Rays using the Principal Axes of the Directional Energy Distribution}

\shorttitle{Detecting Local Deflection Patterns of UHECR using Principal Axes}

\authors{
Martin Erdmann$^{1}$ and Tobias Winchen$^{1}$
}

\afiliations{
$^1$ III. Physikalisches Institut A, RWTH Aachen University\\
}

\email{tobias.winchen@physik.rwth-aachen.de}

\abstract{
	From deflections in galactic and extragalactic magnetic fields energy
	dependent structures in the arrival directions of ultra-high energy
	cosmic rays (UHECR) are expected. We propose to characterize these
	structures by the strength of collimation of energy along the
	principal axes in selected regions in the sky. While the strength of
	collimation are indicators of anisotropy in the arrival distribution
	of UHECR, the orientation of the principal system holds information
	about the direction of the deflections of UHECR. We discuss the method
	and present expected limits on the strength of deflection and density
of sources using simulated scenarios of UHECR proton propagation.}


\begin{document}
\maketitle
\section{Introduction}

Ultra-high energy cosmic rays (UHECR) are likely accelerated in
extragalactic point sources. Identifying such sources has not been
successful so far, presumably due to the deflection of the charged cosmic rays in the galactic and
extragalactic magnetic fields. However, the deflection of UHECR during
their propagation can be quantitatively modelled using simulation
software (e.g.~\cite{Sutherland2010, Aloisio2012c, Kampert2013,
Bretz2013, DeDomenico2013}). To compare the predictions of the models
with the datasets collected by experiments like the Pierre Auger
Observatory~\cite{PAO2010b, Abraham2010} or Telescope
Array~\cite{Abu-Zayyad2012, Tokuno2012}, observables are needed that
discriminate between different astrophysical scenarios.  In particular, analysis of
the energy and arrival directions of UHECR can probe cosmic magnetic
fields and the density of sources of UHECR~\cite{Lee1995,PAO2012,Abreu2013a}.
The expected deflection  patterns  can be  abstracted as symmetric
`blurring' from multiple scattering in turbulent fields and threadlike
structures from deflection in coherent fields.  In a localized region in
the sky, further on denoted as region of interest (ROI), both effects
result in a collimation of energy along the axes of the principal system
of the directional energy distribution.

\section{The Thrust Observables}
To derive the principal axes and quantify the collimation of energy
along these axes, we use here the `thrust observables' that were first used in high
energy physics to characterize the energy distribution in particle
collisions~\cite{Brandt1964}. 
The three thrust observables $T_{k=1,2,3}$ quantify the strength of the
collimation of the particle momenta along each of the three axes
$\vec{n}_{k=1,2,3}$ of the principal system.  The principal axes and the
corresponding observables $T_k$ are successively  determined by
maximizing $T_k$ with respect to the axis $\vec{n}_k$ using
\begin{equation}
	T_k = \max_{\vec{n}_k} \left(\frac{\sum_i |\vec{p}_{i}
	\vec{n}_k|}{\sum_i |\vec{p}_i|} \right)
	\label{eq:ThrustEquation}
\end{equation}
with $\vec{p}_i$ being the momentum  of the individual particles. For $k
= 1$ the quantity $T_1$ is called `thrust' and consequently the first
axis of the principal system $\vec{n}_1$ is called `thrust axis'. For
the second axis the additional side condition $\vec{n}_1 \perp
\vec{n}_2$ is used in eq.~\ref{eq:ThrustEquation}. The resulting
value $T_2$ is denoted as `thrust major', the axis as `thrust major
axis'. Finally, the third quantity $T_3$ is called `thrust minor' with
corresponding `thrust minor axis'.  For the thrust minor axis
$\vec{n}_3$ it is $\vec{n}_1 \perp \vec{n}_2 \perp \vec{n}_3$ which
renders the maximization in eq.~\ref{eq:ThrustEquation} trivial.

To use these observables in astroparticle physics, we calculate them from
the momenta $\vec{p}_i$ of all events in a small circular region of the sky.  As all observed cosmic
rays approach the observer centered in the coordinate system, the thrust
axis points to the barycenter of the energy distribution in this region.
In spherical coordinates, the thrust axis is anti-parallel to the radial unit vector $\vec{e}_r$
pointing to the local barycenter of the energy distribution.  The thrust
major and thrust minor axes can therefore be written as linear
combinations of the unit vectors $\vec{e}_\phi$ and $\vec{e}_\theta$
reading 
\begin{equation} \vec{n}_{2,3} = \cos{\xi_{2,3}} \cdot
	\vec{e}_\phi + \sin{\xi_{2,3}} \cdot \vec{e}_\theta 
\end{equation}
with $\xi_3 = 90^\circ + \xi_2$. Using this together with
eq.~\ref{eq:ThrustEquation}, $T_2$ becomes maximal if $\vec{n}_2$ is
aligned to a linear distribution of UHECR\@. The thrust major axis thus
points along threadlike structures in the energy distribution of UHECR.
As the thrust minor axis $\vec{n}_3$ is chosen perpendicular to
$\vec{n}_1$ and $\vec{n}_2$ its direction has no physical meaning beyond its
connection to the thrust major axis. The corresponding thrust minor
value $T_3$ holds additional information beyond the values of $T_1$ and
$T_2$.

We include all cosmic rays with energy above $E_{\min} = \SI{5}{\EeV}$
in the calculations and set the radius of the circular region of interest to $\beta
= \SI{0.25}{\radian}$. These values have been chosen to maximize the
discriminating power of propagation simulations from isotropic
distributions of UHECR~\cite{Winchen2013}.

\section{Example simulation}
\begin{figure*}[!t]
  \centering
	\includegraphics[width=\textwidth]{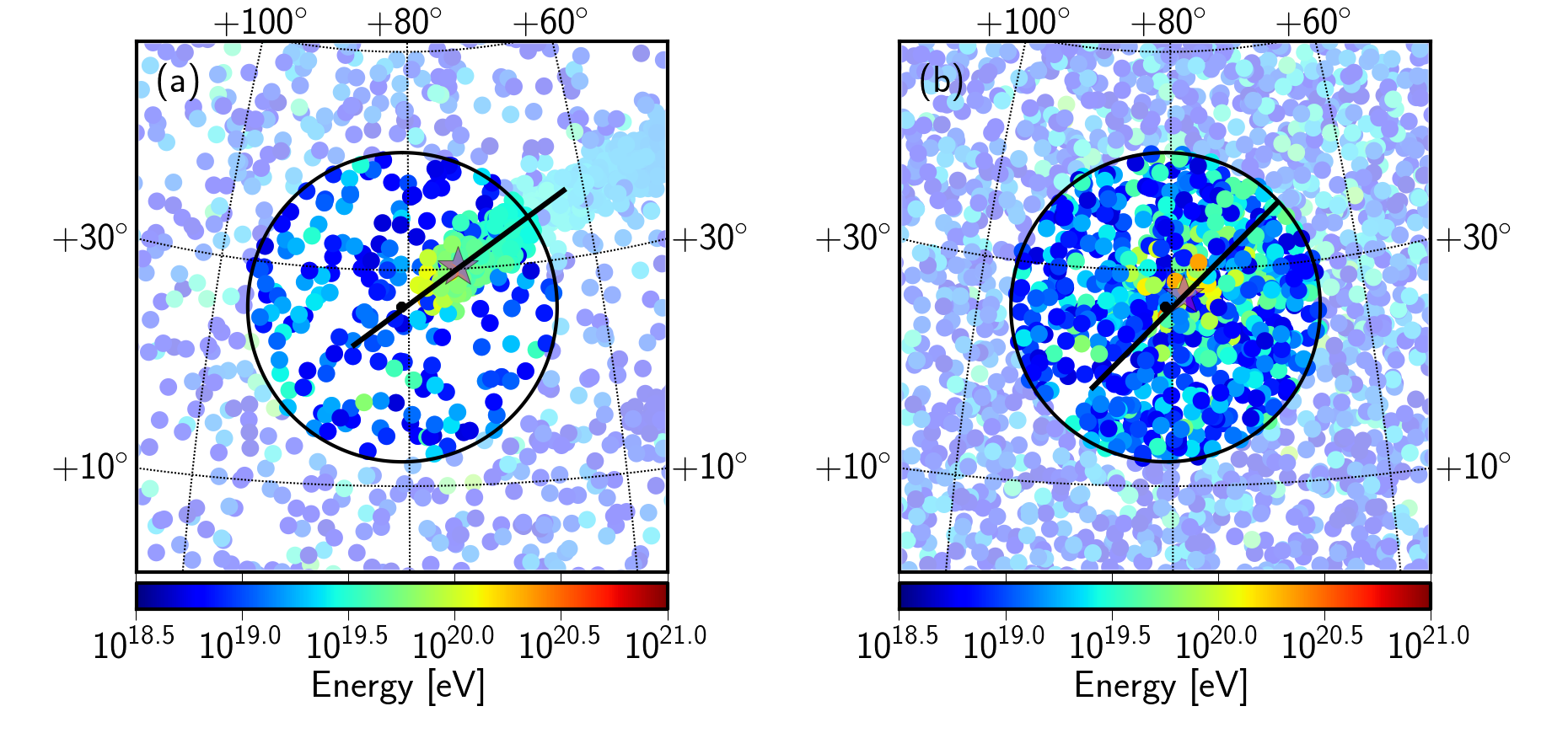}
\caption{Region of interest around the closest source in two 
		simulations with different strength of the EGMF $B =
		\SI{.1}{\nano\gauss}$ (left panel) and $B = \SI{5}{\nano\gauss}$
		(right panel). Colored dots denote arrival direction and energy of
		the UHECR\@. Source position, source density, and galactic magnetic
		field model are identical in both simulations. The thrust axis in
		the regions is denoted by a magenta star; the thrust major axis in
	this region is denoted by a black line.}\label{fig:ParsecExampleClosestSource}
 \end{figure*}
\begin{figure*}[!t]
  \centering
	\includegraphics[width=\textwidth]{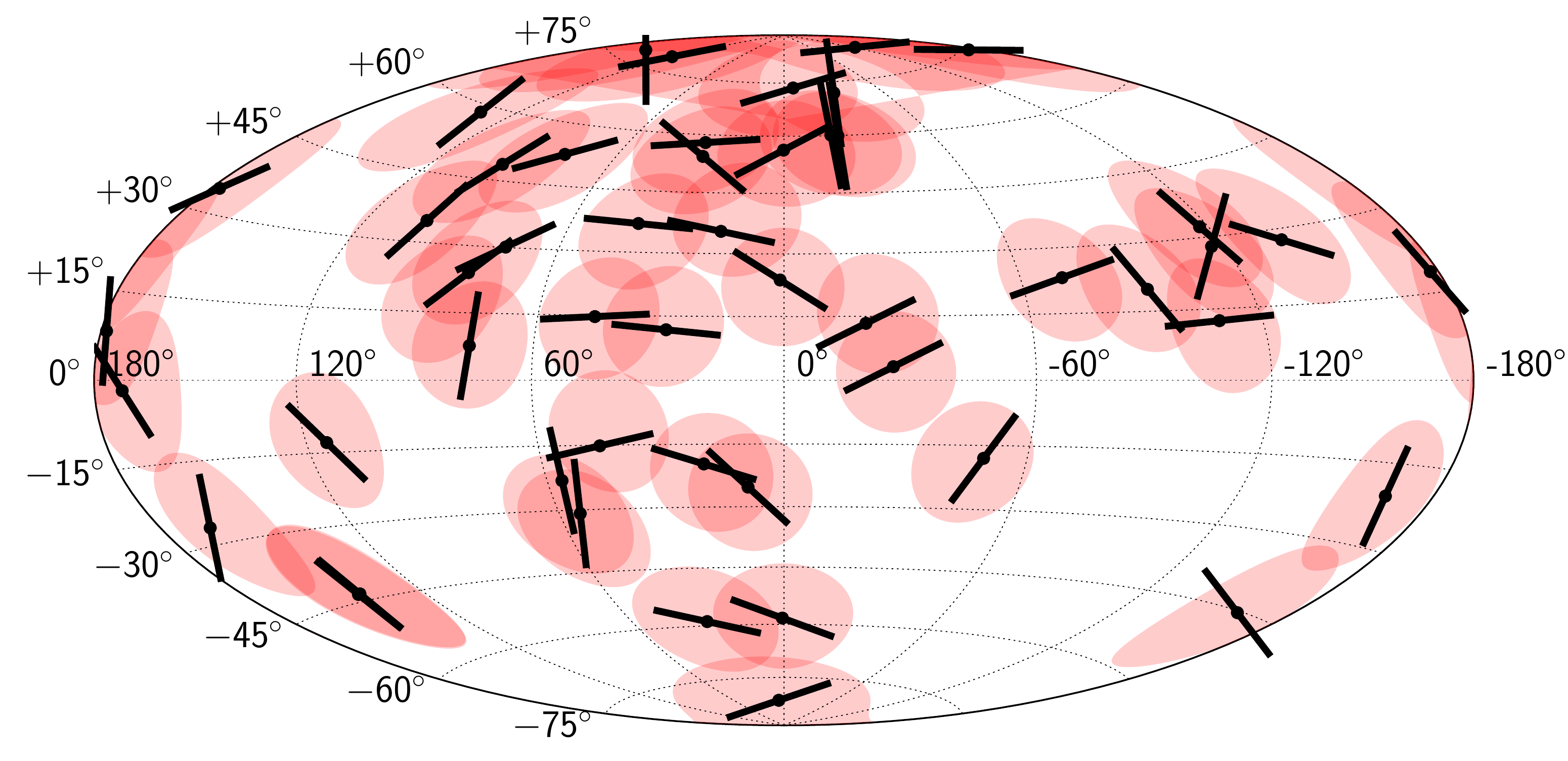}
  \caption{
	Typical skymap in galactic coordinates of thrust major axes (black
	lines) from
		anisotropically distributed UHECR from a PARSEC simulation with
		\SI{0.1}{\nano\gauss}	extragalactic magnetic field, source density
		\SI{d-5}{\per\cubic\mega\parsec}, and JF2012 model for the galactic
	magnetic field. Red shaded areas denote the individual ROI.}\label{fig:ExemplarySkymaps}
 \end{figure*}
To demonstrate the thrust observables, we simulated 20~000 UHECR protons
from homogeneously distributed point sources with a density
\SI{1d-5}{\per\cubic\mega\parsec} in two scenarios with strengths of
the extragalactic magnetic field $B = \SI{.1}{\nano\gauss}$ and $B =
\SI{5}{\nano\gauss}$ using the PARSEC software~\cite{Bretz2013}. The
galactic magnetic field is modeled using a lens for the regular
component of the JF2012~\cite{Jansson2012} magnetic field. The position
of the sources is identical in both simulations. All sources are
simulated with equal luminosity, a power law spectrum with spectral
index $\gamma = -2.7$, and a maximum energy of \SI{1000}{\EeV}.  Regions
of interest with a size $\beta = \SI{0.25}{\radian}$ are set to the
closest 50 sources in the simulations.

In figure~\ref{fig:ParsecExampleClosestSource} the region around the
closest source in the simulations is shown. A magenta star marks the
direction of the thrust axis and a black line denotes the direction of
the thrust major axis.  For the weak extragalactic magnetic field shown
in figure~\ref{fig:ParsecExampleClosestSource}~(a), a tail of UHECR
from the source resulting from coherent deflection is visible. The
thrust major axis points along this structure. Because of the stronger
deflections in the extragalactic magnetic field, the structure is not
visible by eye in figure~\ref{fig:ParsecExampleClosestSource}~(b).
Nevertheless, the thrust major axis points in a similar direction in
this example, indicating the preferred direction of deflection in the
magnetic field.
The values of thrust observable $T_{1}$ calculated in
both cases deviates from the isotropic expectations by more than three
times the spread of the corresponding isotropic distribution.
Note that the observation of a single non-trivial ROI
can be sufficient evidence for an anisotropic arrival distribution of
UHECR.
The complete map of thrust major axes of this example 
is shown in figure~\ref{fig:ExemplarySkymaps}. The map indicates the
deflection patterns of cosmic rays in the magnetic fields.

\begin{figure}[t]
  \centering
	\includegraphics[width=0.5\textwidth]{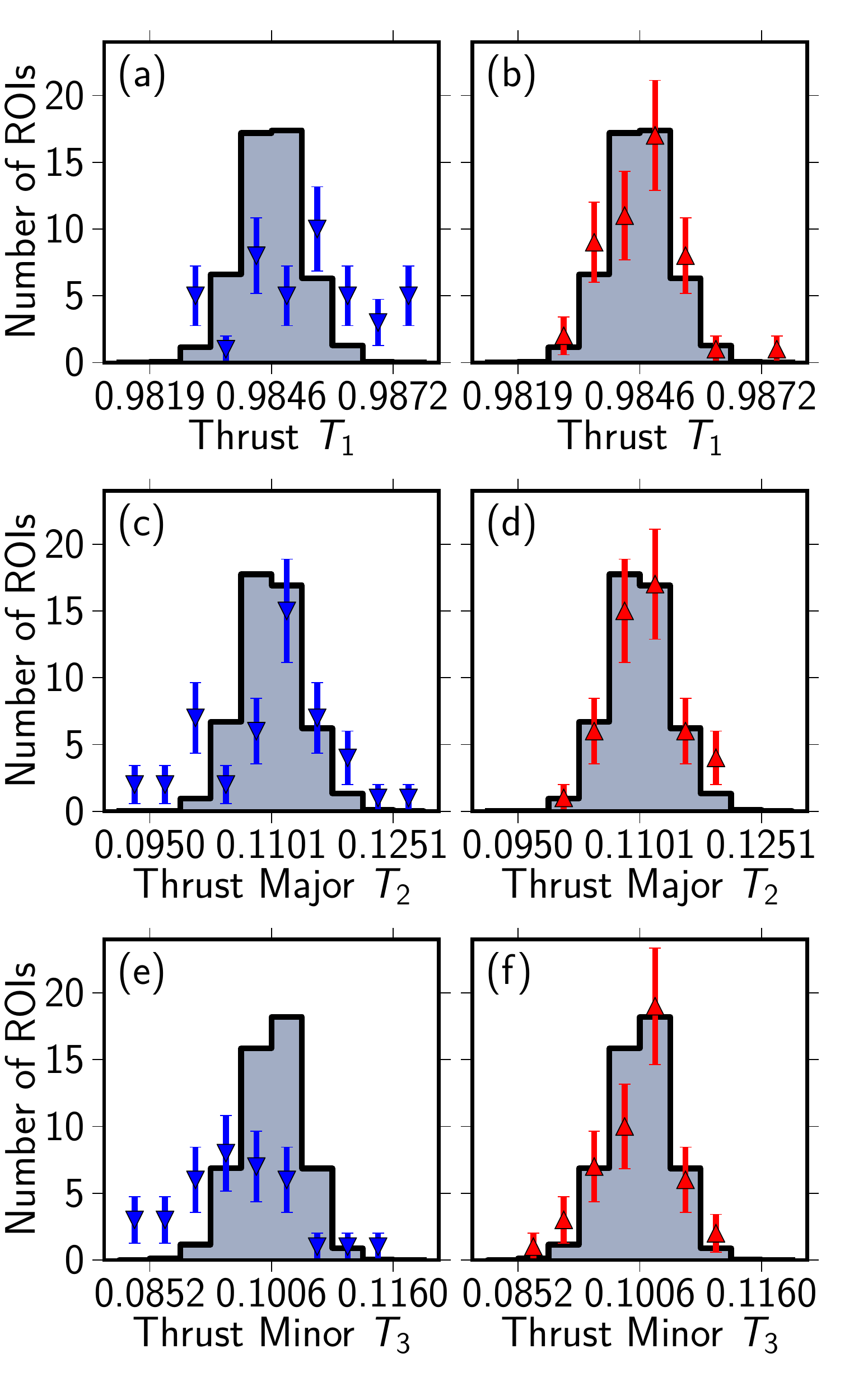}
	\caption{Distribution of observables $T_{1,2,3}$ in an example
	scenario using protons from homogeneously distributed point sources with density $\rho =
	\SI{1d-4}{\mega\parsec^{-3}}$, JF2012 regular galactic magnetic field,  and strength of the extragalactic
	magnetic field \textbf{(a,c,e)}~$B =	\SI{.1}{\nano\gauss}$, and
	\textbf{(b,d,f)}~$B = \SI{5}{\nano\gauss}$. 
	The gray histograms correspond to the
		average of the observables from 100 simulations
		with isotropically distributed UHECR.\@ 
		}\label{fig:ParsecExampleObservableDistribution}
 \end{figure}
In figure~\ref{fig:ParsecExampleObservableDistribution} the
corresponding distributions of the observables $T_{1,2,3}$ are shown
together with the mean of 100 simulations with isotropically distributed
UHECR\@.  For weak extragalactic magnetic fields, the distributions for
$T_{1,2,3}$ deviate considerably in several ROI from the expectation for
isotropically distributed UHECR\@.  For  $B = \SI{5}{\nano\gauss}$, in
this example only the thrust of a single ROI deviates from the isotropic
expectation. 

In the example above, we calculated the observables in regions
centered at the sources of UHECR.\@ As the sources of UHECR are, however,
yet unknown, this is not possible in the analysis of measured data. In
the analysis following  below, we
therefore choose regions around events with energy $E >
\SI{60}{\EeV}$ as ROI, assuming that UHECR with the highest energies
are least deflected. 

\section{Statistical Interpretation}
Searches for anisotropy and structure in the arrival directions of UHECR
did not yet lead to strongly conclusive results.
In order to exploit the sensitivity of our method to
non-trivial astrophysical scenarios, simulated UHECR data sets can be
generated with arbitrary small signal contributions.
By comparison of the simulated results
with the observation, thus limits on the simulated astrophysical model parameters can be
set using these measurements.

For the thrust observables described above,  the likelihood ratio 
\begin{equation}
	Q = -2
	\ln\frac{\mathcal{L}_{\mathcal{H}_\vecN{X}}}{\mathcal{L}_{\mathcal{H}_0}}
\end{equation}
is used as test statistic. The likelihood  
\begin{equation}
	\mathcal{L}_{\mathcal{H}_x} = \prod_i \binom{N}{r_i}
	p_i^{r_i} {(1 - p_i)}^{N-r_i}.
	\label{eq:LikelihoodHistogramm}
\end{equation}
is calculated from the probabilities to observe $r_i$ out of N ROI with
observable value $T_k$ in bin $i$, where $p_i$ is the probability to
observe a ROI in the bin in scenario $\mathcal{H}_\vecN{X}$.

In frequentist interpretation, $P(Q > Q_{\text{obs}} |
\mathcal{H}_\vecN{X})$ is the frequency of occurrence of $Q >
Q_{\text{obs}}$ in repeated experiments if $\mathcal{H}_\vecN{X}$ is
true.  If both
hypotheses are clearly distinguishable in the analysis, $P(Q >
Q_{\text{obs}} | \mathcal{H}_\vecN{X})$ provides a good estimator for
the confidence in the alternative hypothesis.  If, however, the
hypotheses are only marginally distinguishable, a fluctuation of
$Q_{\text{obs}}$ to a large value results in low confidence in the
alternative hypothesis if the confidence is estimated as above.  A derivation of limits
on parameter $\vecN{X}$ with this method thus prematurely excludes
scenarios, to which the analysis is not sensitive.

To avoid this in frequentist inference, a modified likelihood
ratio can be used instead to calculate the confidence in the signal
hypothesis~\cite{Read2000, Read2002}. This
CL\textsubscript{S} method is, e.g., used to identify valid 
mass ranges for the Higgs Boson at the LEP~\cite{Barate2003},
Tevatron~\cite{Aaltonen2010}, and LHC~\cite{Aad2012, Chatrchyan2012}
experiments. Here, the confidence in the signal hypothesis
 $\mathcal{H}_\vecN{X}$ is defined as
\begin{equation}
	CL_S = \frac{P(Q > Q_{\text{obs}} | \mathcal{H}_{\vecN{X}})}
	{P(Q > Q_{\text{obs}} |\mathcal{H}_0)}.
	\label{eq:CLSMethod}
\end{equation}
This corresponds to a weighting of the probability to get
$Q_{\text{obs}}$ if $\mathcal{H}_{\vecN{X}}$ is true with the confidence
in the background-only hypothesis $\mathcal{H}_0$. Points in parameter
space with, e.g., $CL_S < 0.05$ are excluded at 95\% confidence. 

In the PARSEC simulation software used here  deflections in the
extragalactic magnetic field are assumed to be symmetric around the
sources, resulting from long propagation distances through unstructured
magnetic fields. For structured magnetic fields, and also for turbulent
fields with short propagation distances, this overestimates the
deflection strength. As
the extragalactic magnetic field is likely structured, we discuss here
primarily limits on the strength of the deflection $C_{eg}$ with average
deflection 
\begin{equation}
	\delta = C_{eg} \sqrt{\frac{D}{\si{\mega\parsec}}}
	{\left(\frac{E}{\si{\exa\electronvolt}} \right)}^{-1}
	\label{eq:DefinitionOfCT}
\end{equation}
for UHECR with energy $E$ from a source in distance $D$ 
as alternative to limits on the fieldstrength $B$.

\begin{figure*}[!t]
  \centering
	\includegraphics[width=\textwidth]{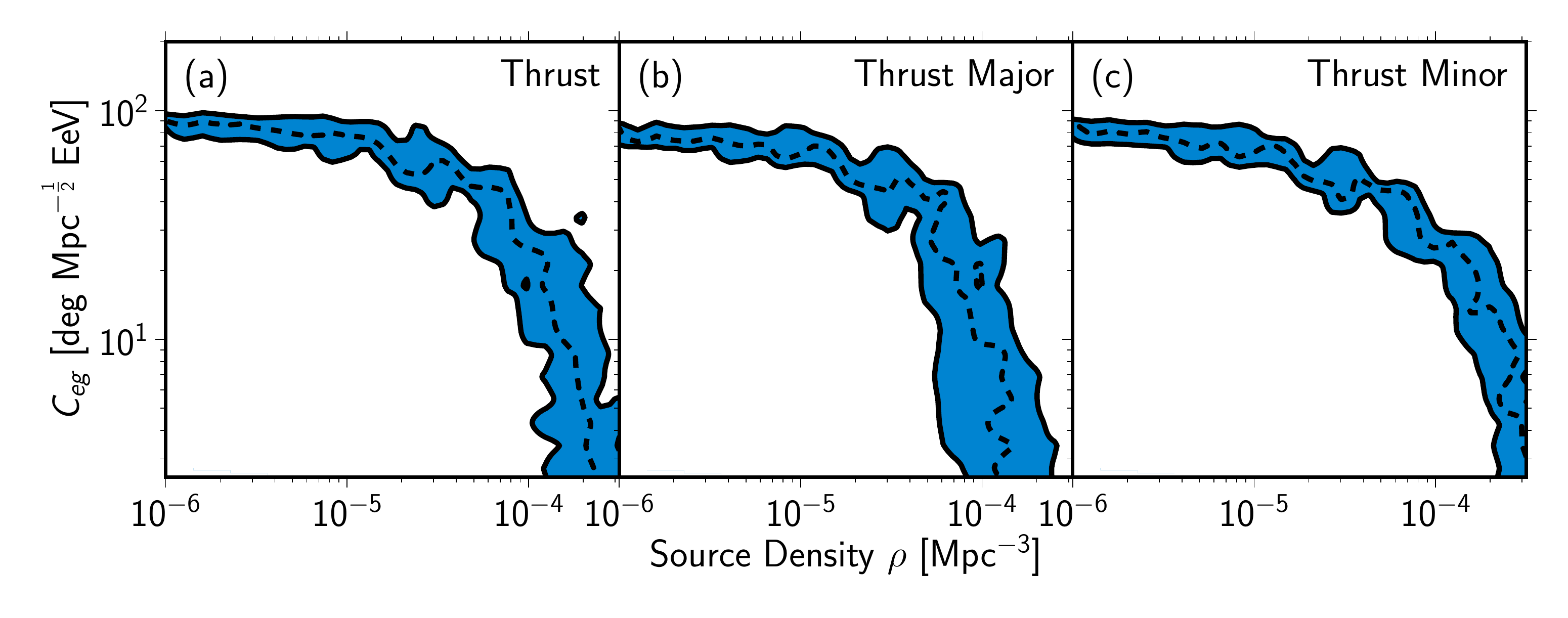}
  	\caption{Expected exclusion limits on the deflection strength in the
		extragalactic magnetic field $C_{eg}$ for $\sim\SI{20000}{}$ observed
		UHECR protons (dashed line) as a function of the source density.  The blue shaded area
		denotes the limits for a $\pm 1\sigma$ fluctuation of the
		observed likelihood ratio $Q_{obs}$ in equation~\ref{eq:CLSMethod}.}\label{fig:ExpectedLimits}
  \label{fig:ExpectedExclusionLimits}
 \end{figure*}
In figure~\ref{fig:ExpectedExclusionLimits} the expected limits on the
deflection strength $C_{eg}$ in the extragalactic magnetic field as a function of the density of point sources
in the simulations is shown for $\sim\SI{20000}{}$ detected UHECR protons above
\SI{5}{\EeV}. In the simulations, deflections expected from a
JF2012~\cite{Jansson2012} regular field and a limited field of view of a
typical earth-bound observatory are included~\cite{Sommers2001}. To account for the non-uniform exposure, the energies of the
individual UHECR in eq.~\ref{eq:ThrustEquation} are weighted by the
relative exposure in the arrival direction.
By measuring the
thrust observables using a current UHECR experiment, scenarios  can be
tested, in which UHECR are protons that originate from point sources with a density
less than $\rho \approx \SI{1d-5}{\per\cubic\mega\parsec}$ and exhibit
deflections weaker
than $C_{eg} \approx \SI{90}{\degree\per\mega\parsec\tothe{1/2}\EeV}$
in the extragalactic magnetic field.

\section{Conclusion}
We presented a
method to characterize the directional energy distribution of UHECR using
the thrust observables from high energy physics. The directions of
preferred deflection are identified as directions of the principal axes
with this method. The distribution of thrust observables measured in
localized region in the sky can be used to compare observations with
predictions from model scenarios. For UHECR being protons, we estimated
that with the statistic of current UHECR experiments, scenarios with
deflections up to $C_{eg} \approx
\SI{90}{\degree\per\mega\parsec\tothe{1/2}\EeV}$ can be tested, if the
density of sources is compatible with the density of radio loud AGN.

\vspace*{0.5cm}
\footnotesize{{\bf Acknowledgment:~}{
This work is supported by the Mi\-nis\-ter\-ium
f\"ur Wis\-sen\-schaft und For\-schung, Nordrhein-Westfalen, and the
Bundes\-mi\-nis\-ter\-ium f\"ur Bil\-dung und For\-schung (BMBF). T.
Winchen gratefully acknowledges funding by the
Friedrich-Ebert-Stif\-tung.}


\begin{thebibliography}{10}

\bibitem{Sutherland2010}
M.~S. Sutherland, B.~M. Baughman, J.~J. Beatty, {\it Astroparticle Physics\/}
  {\bf 34}, 198 (2010).

\bibitem{Aloisio2012c}
R.~Aloisio, D.~Boncioli, A.~Grillo, S.~Petrera, F.~Salamida, {\it JCAP\/} {\bf
  1210}, 007 (2012).

\bibitem{Kampert2013}
K.-H. Kampert, {\it et~al.\/}, {\it Astroparticle Physics\/} {\bf 42}, 41
  (2013).

\bibitem{Bretz2013}
H.-P. Bretz, M.~Erdmann, P.~Schiffer, D.~Walz, T.~Winchen, {\it
	arXiv:1302.3761,	submitted to  {Astroparticle Physics}\/}  (2013).

\bibitem{DeDomenico2013}
M.~De~Domenico, {\it arXiv:1305.4364\/}  (2013).

\bibitem{PAO2010b}
J.~Abraham, {\it et~al.\/}, {\it Nuclear Instruments and Methods\/} {\bf A613},
  29 (2010).

\bibitem{Abraham2010}
J.~Abraham, {\it et~al.\/}, {\it Nuclear Instruments and Methods} {\bf 620}, 227 (2010).

\bibitem{Abu-Zayyad2012}
T.~Abu-Zayyad, {\it et~al.\/}, {\it Nuclear Instruments and Methods} {\bf A689}, 87
  (2012).

\bibitem{Tokuno2012}
H.~Tokuno, {\it et~al.\/}, {\it Nuclear Instruments and Methods} {\bf A676}, 54 (2012).

\bibitem{Lee1995}
S.~Lee, A.~V. Olinto, G.~Sigl, {\it Astrophysical Journal\/} {\bf 455},
  L21–L24 (1995).

	\bibitem{PAO2012}
P.~Abreu, {\it et~al.\/}, {\it Astroparticle Physics\/} {\bf 35}, 354 (2012).

\bibitem{Abreu2013a}
P.~Abreu, {\it et~al.\/}, {\it JCAP (in press), arXiv:1305.1576} (2013).


\bibitem{Brandt1964}
S.~Brandt, C.~Peyrou, R.~Sosnowski, A.~Wroblewski, {\it Physics Letters\/} {\bf
  12}, 57 (1964).

\bibitem{Winchen2013}
T.~Winchen, Ph.D. thesis, RWTH Aachen University (Submitted in April 2013).

\bibitem{Jansson2012}
R.~Jansson, G.~R. Farrar, {\it Astrophysical Journal\/} {\bf 757}, 14 (2012).

\bibitem{Read2000}
A.~L. Read, {\it Proceedings of the 1st Workshop on Confidence
Limits\/}, 81, (CERN, Geneva, 2000).

\bibitem{Read2002}
A.~L. Read, {\it Journal of Physics G\/} {\bf 28}, 2693 (2002).

\bibitem{Barate2003}
R.~Barate, {\it et~al.\/}, {\it Physics Letters B\/} {\bf 565}, 61 (2003).

\bibitem{Aaltonen2010}
T.~Aaltonen, {\it et~al.\/}, {\it Physical Review Letters\/} {\bf 104}, 061802
  (2010).

\bibitem{Aad2012}
G.~Aad, {\it et~al.\/}, {\it Physical Review D\/} {\bf 86}, 032003 (2012).

\bibitem{Chatrchyan2012}
S.~Chatrchyan, {\it et~al.\/}, {\it Physics Letters B\/} {\bf 710}, 26 (2012).

\bibitem{Sommers2001}
P.~Sommers, {\it Astroparticle Physics\/} {\bf 14}, 271 (2001).

\end{thebibliography}
\end{document}